# Photonic Temperature and Wavelength Metrology by Spectral Pattern Recognition

SIEGFRIED JANZ[1,*], ROSS CHERITON[1,2], DAN-XIA XU[1], ADAM DENSMORE[2], SERGEY DEDYULIN[3], ANDREW TODD[3], JENS H. SCHMID[1], PAVEL CHEBEN[1], MARTIN VACHON[1], MOHSEN KAMANDAR DEZFOULI[1], DANIELE MELATI[1]

[1]*Advanced Electronics and Photonics Research Centre, National Research Council Canada, Building M-50, 1200 Montreal Rd. Ottawa, ON, Canada, K1A 0R6*
[2]*Herzberg Astronomy and Astrophysics Research Centre, National Research Council Canada, 5071 West Saanich Road, Victoria, BC, Canada, V9E 2E7*
[3]*Metrology Research Centre, National Research Council Canada, Building M-36, 1200 Montreal Rd. Ottawa, ON, Canada, K1A 0R6*

*\*Siegfried.janz@nrc-cnrc.gc.ca*

**Abstract:** Spectral pattern recognition is used to measure temperature and generate calibrated wavelength/frequency combs using a single silicon waveguide ring resonator. The ring generates two incommensurate interleaving TE and TM spectral combs that shift independently with temperature to create a spectral pattern that is unique at every temperature. Following an initial calibration, the ring temperature can be determined by recognizing the spectral resonance pattern, and as a consequence the wavelength of every resonance is also known. Two methods of pattern based temperature retrieval are presented. In the first method, the ring is locked to a previously determined temperature set-point defined by the coincidence of only two specific TE and TM cavity modes. Based on a prior calibration at the set-point, the ring temperature and hence all resonance wavelengths are then known and the resulting comb can be used as a wavelength calibration reference. In this configuration, all reference comb wavelengths have been reproduced within a 5 pm accuracy across an 80 nm range by using an on-chip micro-heater to tune the ring. For more general photonic thermometry, a spectral correlation algorithm is developed to recognize a resonance pattern across a 30 nm wide spectral window and thereby determine ring temperature continuously to 50 mK accuracy. The correlation method is extended to simultaneously determine temperature and to identify and correct for wavelength calibration errors in the interrogating light source. The temperature and comb wavelength accuracy is limited primarily by the linewidth of the ring resonances, with accuracy and resolution scaling with the ring quality factor.

## 1. Introduction

Silicon photonic ring resonators are integrated optical filters with a transmission spectrum consisting of a comb of resonance peaks with a quasi-periodic wavelength spacing, or free spectral range (FSR), determined by the ring path length $L$ and wavelength-dependent effective index $N_{eff}$ [1,2]. These resonances shift with temperature [3,4] because of the temperature dependence of the silicon and waveguide cladding refractive indices [5]. Si ring resonators may therefore be used as optically interrogated thermometers [3,4,6,7]. For example, the temperature metrology community is exploring ring resonator thermometers as a potential replacement [7,8] for standard platinum resistance thermometers (SPRTs). A silicon ring-resonator can offer several advantages over a SPRT. These include a much better immunity to mechanical shock and chemical contamination (and hence less frequent re-calibrations) [9,10], and also an alternative traceability chain that is based on optical frequency rather than electrical resistance. The Si sensor element is also very small, with typical diameters of 100 μm or less as compared to several centimeters for a coiled SPRT wire. Such ring resonators are therefore a better approximation to a single point sensor so that readings are less prone to ambiguity arising from thermal gradients across the thermometer. After initial calibration, the ring temperature can be determined with accuracies well below 1 K [6-8] by tracking the changing resonance wavelengths.

Ring resonators can also be used to provide optical frequency (or wavelength) combs for calibrating spectroscopic measurements and optical instrumentation. In fields such as astronomy that rely on precise spectroscopic measurements (e.g., for stellar radial velocity and proper motion measurements), spectrometers must be calibrated at the time of measurement using spectral reference sources. The most common method is to use atomic or molecular emission lines from reference lamps, but these do not always provide spectral lines at the desired wavelength range or line intervals [11,12]. Frequency combs generated by nonlinear optical mechanisms and mode-locked lasers can provide broad regularly spaced octave spanning frequency combs that can be directly referenced to atomic transitions [13-15]. Such systems provide the ultimate in achievable accuracy, but are complex and not easily portable. The mode frequency spacing set by the associated mode-locked lasers of a few hundred MHz may also be smaller than desired for many spectroscopy applications. A much simpler passive Fabry-Perot cavity [16] or its integrated optical equivalent, the waveguide ring resonator [17], can be used to generate a frequency comb for calibration purposes, provided that the cavity is stabilized to a sufficient accuracy. Comb frequency spacings of a few GHz (approx. 10 pm (0.1 Å) in wavelength) to several hundred GHz (several nm) are easily produced by choosing an appropriate ring resonator cavity length.

The use of ring resonators for either temperature measurement or wavelength calibration does involve inter-related implementation challenges. If a ring resonator is used to provide a wavelength comb for spectral calibration purposes, the temperature of the resonator must be known with sufficient precision that the resonant wavelengths fall within the required tolerances. For silicon ring resonators, a temperature uncertainty of 1 °C corresponds to a resonance wavelength uncertainty approaching 0.1 nm (1Å). On the other hand, when using ring resonators for thermometry [3,6-8], the temperature accuracy depends on the calibration stability of the light source or spectrometer used to follow the resonance wavelengths. Ring resonator sensors also have a cyclic response to temperature, as do most optical interferometric sensors. The operating range of a ring thermometer is thus limited to the temperature interval over which a resonance shifts by one FSR, since the resonances of a single ring can be almost indistinguishable from each other. An FSR large enough to accommodate a temperature range of more than 100 K requires rings with radii approaching 10 μm, below which bend losses become significant. For larger rings with small FSR [7], temperature can be monitored over a wide range by tracking a selected resonance continuously as temperature evolves, provided the measurement is referred to an initial baseline measurement at a known starting temperature. The above considerations to some extent negate the advantages of small size and simplicity that photonics may provide. In this paper we propose a transduction method by which the temperature of the ring is uniquely determined by identifying a spectral resonance pattern at any instant in time, with no need to keep continuous track of resonances. There is no intrinsic limit on operating range provided the pattern is not cyclic. Finally, since temperature is encoded in a pattern rather than any specific resonance wavelength, the temperature reading accuracy is more tolerant to wavelength calibration errors in the interrogating optical system. Pattern recognition is to some extent self-calibrating, and the precision requirements for the associated instrumentation can be considerably relaxed. Such a device can be a useful standard for both temperature and frequency/wavelength metrology.

In order to generate a suitable temperature dependent pattern, one may consider two ring resonators having incommensurate ring lengths and different thermo-optic responses. Such a pair of rings will have transmission comb spectra with a different FSR and the resonances will shift at different rates with temperature. Two rings with the desired differences in optical properties can be fabricated by using different silicon waveguide dimensions [4], or subwavelength structures, possibly in combination with different cladding materials [18]. Cascading two such dissimilar rings has been used to generate, through the Vernier effect, a comb with an intensity envelope that shifts much more rapidly with temperature than the individual resonances [4]. While the Vernier approach increases the sensitivity of a ring based thermometer, the Vernier envelope response is itself cyclic so the temperature range is limited to one Vernier period. We focus instead on the underlying pattern of interleaving resonances in a two resonator system. When considered together, two incommensurate ring spectra comprise a pattern that is unique for any given chip temperature. In this paper, rather than two separate rings, we use a single ring resonator that supports both transverse electric (TE) waveguide modes and transverse magnetic (TM) waveguide modes. Since the modes are orthogonal the transmission spectra of the two polarization modes can be considered to arise from independent resonator cavities. Light in both modes is transmitted through the same ring simultaneously, and the output signal is simply a linear superposition of the two spectra. This configuration simplifies data acquisition since only one input and output is required, and also eliminates any effect of temperature gradients across the chip that would occur for spatially separate rings. We show that in this configuration, the ring temperature can be determined from a single measurement of the spectral pattern. Once the spectral patterns have been calibrated against known temperature and wavelength standards, subsequent recalibrations are not required. Furthermore, once the ring temperature is known, the ring can provide a reproducible wavelength comb for spectral calibration purposes.

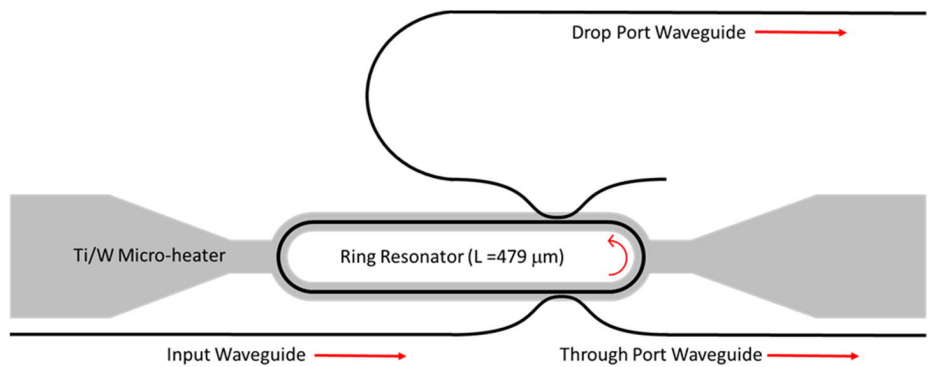

Fig. 1. Schematic layout of silicon ring resonator with an integrated Ti/W heater. The Si waveguides are 450 nm wide and 220 nm thick. The ring round trip length is L= 479 μm.

## 2. Experiment

The silicon-on-insulator (SOI) ring resonator shown in Figure 1 is a 479 μm long waveguide racetrack formed by a 220 nm thick and 450 nm wide Si channel waveguide on a 2 μm buried $SiO_2$ layer with a 2.2 μm $SiO_2$ upper cladding. A measured waveguide loss of approximately 1.3 dB/cm was derived by comparing the insertion loss for several waveguides ranging from 0.8 to 4.8 cm in length. The ring is connected to drop and through port waveguides by directional couplers with a nominal gap of 200 nm. The ring resonator supports both TE and TM polarized resonance modes. An integrated titanium/tungsten resistive metal micro-heater with 200 nm metal thickness was placed over the ring, to allow the ring temperature to be tuned independently of the underlying heated stage. The temperature tuning rate is approximately 250 mK/mA for currents in the 15 to 30 mA range. Light was coupled to and from the chip through polarization maintaining fibers with lensed tips. The input and output Si waveguides were terminated at the chip facet by inverse tapered couplers using subwavelength patterning [19] to adiabatically expand the waveguide mode to match the mode of the lensed fiber.

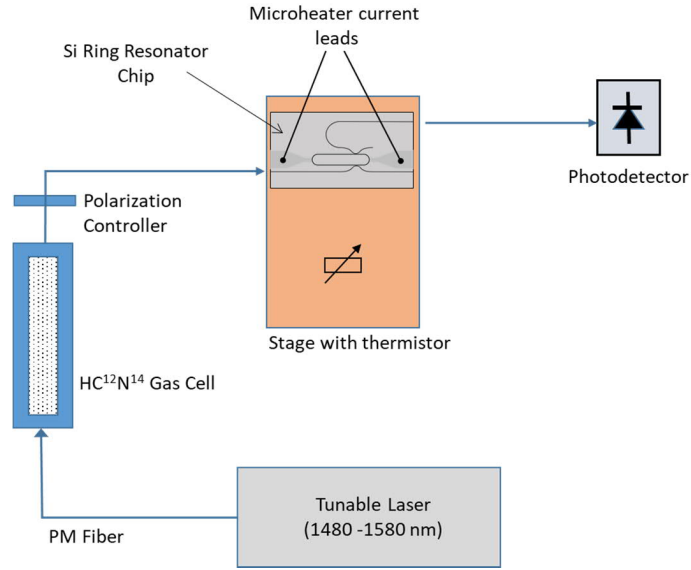

Fig. 2. The experimental configuration for measuring ring resonator spectra and controlling ring temperature using the heated stage and the on-chip micro-heater.

The experimental configuration is shown in Figure 2. All measurements in this work were carried out with the silicon waveguide chip placed on a temperature controlled stage. Stage temperature was measured using a thermistor embedded in the stage. The light source was a fiber coupled tunable laser with wavelength operating range from $\lambda=1480$ nm to $\lambda=1580$ nm. Ring spectra could be acquired by scanning the laser in wavelength steps ranging between 1 pm and 10 pm, depending on the experiment. In these experiments and analysis the temperature and wavelengths given are all referred to the stage thermistor and tunable laser wavelength readings, respectively. With the ring micro-heater off, a one-dimensional heat flow model [20] that includes thermal radiation effects predicts that the difference between stage temperature and waveguide temperature will be less than 50 mK, for an ambient room temperature near 20 °C and stage temperatures between 20 °C to 50 °C as in these experiments. The typical daily fluctuations in laboratory ambient temperature of up to 2 °C should result in corresponding waveguide temperature variations of less than 5 mK relative to the thermistor readings. A comparison of the tunable laser wavelength readings with the P-branch and R-branch absorption lines of a fiber coupled $HC^{12}N^{14}$ gas cell connected in series with the laser and ring confirmed that the laser wavelength settings were repeatable to within 2 pm. Although the integrated optic test system used here is not designed for rigorous metrology studies, the reproducibility of the temperature and wavelength measurements is sufficient for this demonstration of pattern recognition based transduction.

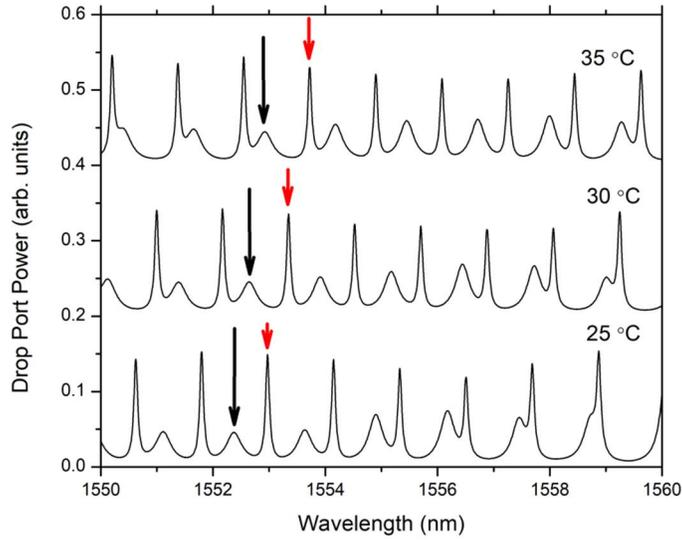

Fig. 3. Measured TE and TM drop port spectra at temperatures of 25, 30 and 35 °C, showing overlapped TE (large peaks) and TM (small peaks) spectra. The arrows indicate the changing positions of a selected pair of adjacent TE (red arrow) and TM (black arrow) resonance peaks at each temperature.

Using the fiber-coupled polarization controller on the input side, the relative power of TE and TM light launched into the chip could be varied. Figure 3 shows the ring drop port spectrum for three temperatures with the TM to TE powers set to an approximate 1:2 ratio. The difference in observed TE and TM resonance linewidths is primarily due to the different ring coupler splitting ratios for the two polarizations. The TM mode extends further into the $SiO_2$ cladding than the TE mode, resulting in an effective index that is much smaller than the TE effective index. At $\lambda = 1550$ nm, the calculated TM and TE effective indices are $N_{eff,TM} \approx 1.73$ and $N_{eff,TE} \approx 2.35$, respectively. Therefore the effective ring cavity optical path length $L_o = L \cdot N_{eff}$ at any wavelength and also the FSR are different for the two polarizations. Since the

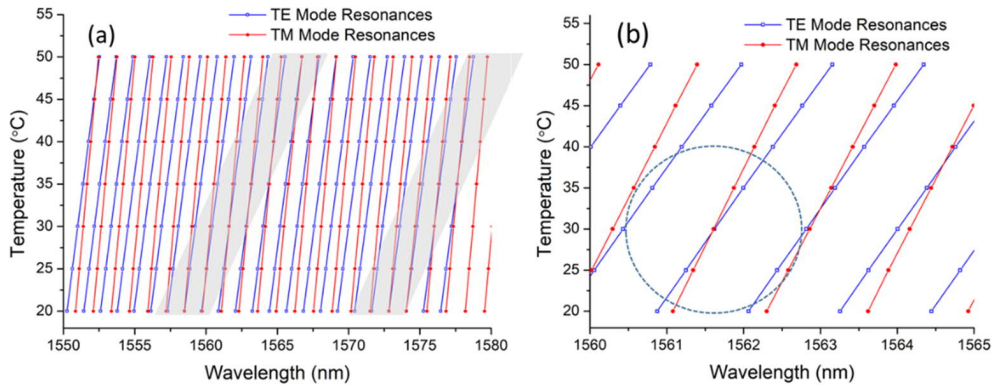

Fig. 4. (a) A temperature-wavelength map of measured TE and TM mode ring resonances between 1550 nm and 1580 and temperatures from 20 °C and 50 °C. Solid lines are a guide to the eye. In the shaded regions the TE and TM resonances are too close to identify separate peaks, and the wavelengths shown are determined by interpolation of nearby measured peak positions. (b) Expanded view of the TE and TM mode map in (a). The dashed circle encloses the coincidence set-point point at which the TE and TM resonances $m_{TE} = 719$ and $m_{TM} = 526$ overlap.

SiO$_2$ cladding thermo-optic coefficient is an order of magnitude smaller than that of Si, the thermo-optic resonance shift for the TM resonances is also smaller than for the TE mode. Figure 4 shows maps of measured TE and TM resonance wavelengths from λ=1550 nm to 1580 nm, over a temperature range from T=20 °C to 50 °C. In Fig. 4 and throughout this work the resonance wavelengths were determined from the peak maxima, as determined by applying a Savitsky–Golay fit in the vicinity of the peak. To a first approximation, the observed resonance wavelengths in Fig. 4 shift linearly with temperature at a rate that is almost independent of wavelength over this range [5]. Approximate shifts of 76 pm/°C and 55 pm/°C for the TE and TM resonances, respectively, are obtained from the data in Figure 4. The important point is that the resonance pattern at any temperature never repeats itself, since the TE and TM free spectral range and thermo-optic shifts are incommensurate. The FSR of the TE and TM modes also varies at different rates with temperature. The temperature of the ring can therefore be determined by recognition of the spectral patterns such as those in Figures 3 and 4, without the need to track specific resonances continuously over time. Once temperature is known from the pattern, then the exact wavelength of every resonance is also known based on initial resonance calibration map such as that given in Figure 4. These comb lines therefore provide a wavelength calibration reference, even in the absence of *a priori* knowledge of the chip temperature and light source wavelength.

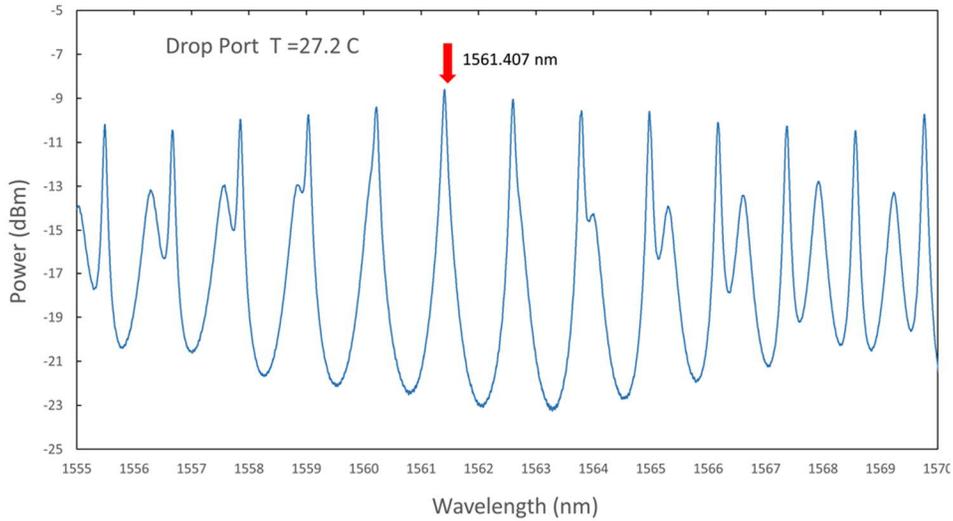

Fig. 5. The measured ring drop port spectrum at the coincidence condition for TE and TM modes of order m$_{TE}$ =719 and m$_{TM}$ = 526, occurring at λ = 1561.407 nm and a stage temperature set at T = 27.2 °C.

## 3. Wavelength and temperature retrieval by mode coincidence set points

The first example of using spectral pattern transduction is to generate a wavelength reference comb using the coincidence of just two resonances, the simplest possible form of spectral correlation. The overlap of any two specific TE and TM resonances defines a set point in a two dimensional (λ, T) plane that is somewhat analogous to the water triple point that occurs at a unique temperature and pressure, and which is used as a convenient reference point for temperature metrology. For example, in Figure 4 the m$_{TE}$ = 719 and m$_{TM}$ = 526 order modes can only coincide at temperature T=27.2 °C and wavelength λ=1561.4 nm. Fig. 5 shows the drop port spectrum observed at this coincidence. The resonance orders m$_{TE}$ and m$_{TM}$ can only be estimated to within about ±2 based on mode solver calculations of $N_{eff}$, due to unavoidable dimensional and compositional differences in the fabricated waveguide from with the nominal design. However this small uncertainty does not significantly alter the results presented here since transduction relies on coincidences and patterns of resonances rather than specific resonance wavelengths.

In order to use the ring to provide a calibrated wavelength comb, all the ring resonance wavelengths would first be calibrated against a known wavelength reference while the ring is maintained at the mode coincidence temperature. Subsequently this calibrated wavelength comb can be reproduced whenever needed simply by tuning the ring temperature to the mode coincidence set-point. Since the set-point is identified from the alignment between the selected TE and TM resonances alone, neither the ring temperature or resonance wavelengths need to be precisely measured to find the set-point. As long as the two modes can be identified, simple feedback signals such as the sum of the coincident TE-TM peak amplitude and/or the peak symmetry at coincidence (accessed using differential frequency and phase sensitive lock-in detection) can be used to tune and hold ring at the set-point temperature. In practice if the nominal incident light wavelength and temperature fall within a few nanometers and several degrees °C of the overlap point (e.g. within the dashed circle of Fig. 4(b)), the identity of the coincident modes is unambiguous.

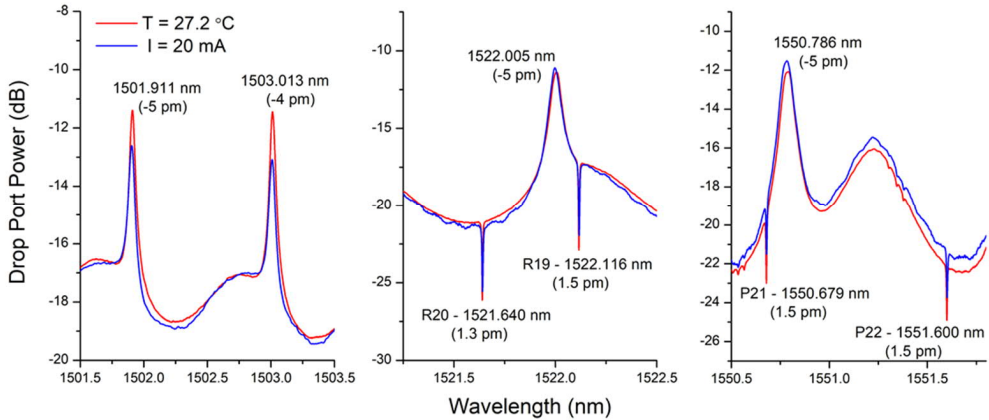

Fig. 6. A comparison of the initial ring spectrum taken at a stage temperature of T= 27.2 °C and heater off (red line), corresponding to the mode coincidence as in Fig. 5, and the recovered spectrum (blue line) when the ring was tuned to the coincidence set point (I =20 mA) by using the on-chip microheater alone. The narrow absorption features are P-branch and R-branch lines of the $HC^{12}N^{14}$ gas cell. The peak offsets between the calibration and recovered spectra are shown in parentheses.

To demonstrate this approach, we recover a calibrated wavelength comb with no knowledge of the chip temperature, using the on-chip micro-heater to tune the ring to the set point defined by the $m_{TE} = 719$ and $m_{TM} = 526$ mode coincidence shown in Fig. 4. The chip was initially calibrated by tuning the ring to the coincidence temperature using the stage heater. The exact coincidence temperature could be determined within 100 mK (0.1 °C) based on the amplitude and symmetry of the overlapped peaks as in Figure 5. These ring resonators were not designed for narrow linewidth, and hence the primary limitation on the accuracy of the set-point arises from the resonance width (approximately 80 pm full width at half maximum (FWHM) for the TE mode, and 400 pm for the TM modes). To recover the calibrated comb spectra, the stage temperature was fixed at 22 °C and the ring micro-heater current alone was used to tune the ring to the mode coincidence reference point. The optimal mode coincidence shown in Figure 5 was found for a current of I=20 mA, as determined by the amplitude and shape of the overlapped resonance feature near $\lambda$=1562 nm. The accuracy with which the set point current can be determined is again limited by the width of the resonances of this ring, and corresponds to a current uncertainty of ±0.5 mA or temperature uncertainty of ±100 mK, similar to the uncertainty in the original set point calibration using the stage heater and thermistor. The full ring spectrum and superposed $HC^{12}N^{14}$ reference spectrum were again recorded at this current. The measured wavelength of the coincident resonance peak at I=20 mA is 1561.402 nm, which is about 5 pm less than the initial measurement using the stage heater. Figure 6 shows a comparison of the calibration spectrum at a stage temperature T=27.2 °C (red lines) and the spectrum recovered using the on-chip heater (blue lines) at three wavelength intervals between 1501 nm and 1552 nm. The spectra in Figure 6 also feature the absorption lines from the $HC^{12}N^{14}$ gas cell in series with the ring. In each spectral interval shown in Fig, 6 the recovered comb lines are all displaced by approximately -5 pm relative to the initial calibration spectrum, while the nearby $HC^{12}N^{14}$ absorption lines confirm that the tunable laser source wavelength reading is repeatable to within 2 pm between runs. By using the correlation method developed in the next section to reassess the ring temperature of both the reference and recovered comb spectra used here, we find that the calibration spectrum yields a temperature of 27.18 (0.05) °C while the recovered spectrum at I = 20 mA gives a temperature of 27.13 (0.05) °C. Although the 50 mK difference between these values is near the confidence limit for correlation temperature determination with this ring design, the difference can account for the -5 pm wavelength offset in the calibration and tuned spectrum in Fig.6.

This first experiment shows that a known calibrated temperature point and calibrated wavelength comb can be recovered on demand by tuning the ring resonator temperature to a set point defined only by the coincidence of two specific TE and TM mode resonances. The primary limitation on comb accuracy is the linewidth of the resonances which limits the resolution in establishing the precise mode coincidence. For a future optimized ring design, the linewidths can be reduced by an order of magnitude by using smaller waveguide-to-ring coupling strengths.

## 4. Temperature and wavelength measurement using spectral correlation

The temperature of the ring is encoded in a continuous manner by the evolving pattern of resonances illustrated Figures 3 and 4, and there is one-to-one correspondence of the spectral pattern to each temperature. Specific mode coincidences as discussed in the previous section can only be used to provide discrete calibrated fixed points in (λ,T)-space at which TE and TM resonance paths cross. However by recording a spectral band encompassing several interleaved TE and TM ring resonances, the ring temperature can be measured over a continuous range.

We use a simple correlation function to implement a spectral pattern recognition algorithm, although there are many possible approaches to pattern recognition including the use of machine learning for more complex systems. Spectral correlation is well established in astronomy [21,22] as a means of detecting gases and extracting precise Doppler shifts from complex stellar and exoplanet spectra. Here a similar methodology is used. Based on a set of initial calibration measurements of the ring spectra at several known temperatures (i.e. those used to create the maps in Figure 4), a temperature dependent reference spectrum $S_R(T)$ can be created based on a look-up table or a parameterized numerical model. The degree of correspondence of the reference spectrum $S_R(T)$ at a trial temperature T with the measured spectrum $S_M$ is evaluated using the correlation function:

$$G(T) = A \left| \int_{\lambda_1}^{\lambda_2} (S_M - \langle S_M \rangle)(S_R(T) - \langle S_R(T) \rangle) \, d\lambda \right|^2 . \qquad (1)$$

Here $\langle S_M \rangle$ and $\langle S_R(T) \rangle$ are the mean values of the measured and reference spectra over the wavelength range from $\lambda_1$ to $\lambda_2$, and $A = [G(S_M, S_M) \, G(S_R(T), S_R(T))]$ is a normalization factor such that $G(T) = 1$ when the measured and reference spectra are identical. The two functions $G(S_M, S_M)$ and $G(S_R(T), S_R(T))$ are the self-correlation functions of the measured and reference spectra respectively. For any given measured spectrum, the corresponding retrieved ring temperature $T_M$ will be the trial temperature T for which the $G(T)$ is a maximum, within a trial temperature interval sufficiently wide to encompass the expected range of $T_M$.

In this work, the model reference spectrum $S_R(T)$ was constructed using polynomial expressions for the waveguide effective indices $N_{eff,TE}$ and $N_{eff,TM}$ derived from the measured TE and TM resonance wavelengths between $\lambda$ = 1550 and 1580 nm and from T = 20 °C to 50 °C. The relation $\lambda_m = N_{eff} L/m$, where $L$ is the ring cavity length and $m$ the resonance order [1,2], was used to calculate the effective indices for all the measured resonance wavelengths. The polynomial expansions in T (°C) and $\lambda$ (μm) for TE and TM modes

$$N_{eff,TE}(\lambda, T) = (2.35243 + 1.97 \times 10^{-4} \cdot T + 1.9 \times 10^{-7} \cdot T^2)$$
$$-(1.2417 + 6.2 \times 10^{-5} \cdot T) \cdot (\lambda - 1.55)) \qquad (2)$$

$$N_{eff,TM}(\lambda, T) = (1.72742 + 1.34 \times 10^{-4} \cdot T + 1.3 \times 10^{-7} \cdot T^2)$$
$$-(1.456 + 3.5 \times 10^{-4} \cdot T) \cdot (\lambda - 1.55)) + 1.64 \cdot (\lambda - 1.55)^2 \qquad (3)$$

were then found by fitting to these experimental $N_{eff}$ values, a. Only terms involving powers of $\lambda$ and T that clearly improved the $N_{eff}$ fitting residuals were retained, and the coefficients in Equations (2) and (3) are shown to the last significant digit that is larger than the uncertainty of the fit. For the final polynomials in Equation (2) and (3) the residual distribution range for calculated and measured resonance $N_{eff}$ values was approximately $10^{-5}$. Equations 2 and 3 were then used in the correlation algorithm to calculate a model $N_{eff}$ and thereby the resonance wavelengths. Once the resonance wavelengths were calculated for a given temperature, the reference spectrum $S_R(T)$ was generated using the ring transmission equations found in Ref. [1]. The histogram distribution of the differences between the calculated reference spectrum resonances and measured resonance wavelengths shown in Fig. 7 has a standard deviation of approximately 4 pm.

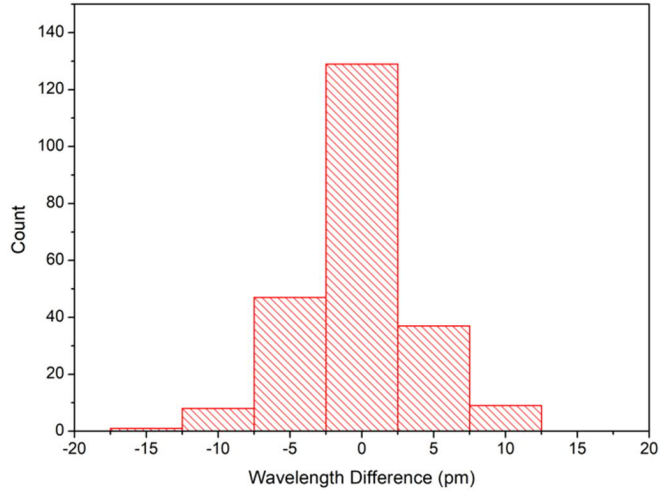

Fig. 7. The distribution of differences between predicted model and measured ring resonance peak wavelengths from $\lambda$ = 1550 to 1580 nm and from T = 20 to 50 °C.

Although temperature is primarily encoded in the pattern of resonance wavelengths, the details of the line shapes and amplitudes of reference spectra can cause small but noticeable variations in the temperature of the correlation function $G(T)$ maximum. The experimental spectral features can be qualitatively reproduced by assuming waveguide-to-ring directional coupler ratios of 0.8/0.2 and 0.55/0.45 for the TE and TM waveguide modes and a TE/TM amplitude ratio of 2:1 when calculating the reference spectrum $S_R$ (T), as shown in Fig. 8. However, we did not attempt to include a detailed wavelength or temperature dependent coupler model in the correlation algorithm, in part because of the sensitivity of the coupling ratios to small variations in the fabricated coupler dimensions that are difficult to characterize. Furthermore, the wavelength dependence of the entire measurement system transfer function, and the interaction of the reference spectrum with the edges of the correlation window (at $\lambda_1$ and $\lambda_2$ in Equation (1)) also cause small shifts in correlation maximum. For metrology applications where much higher accuracy and resolution are required, these effects are best mitigated simply by using ring resonators with much narrower resonance linewidths. The skewing effect of line shape and optical system transfer function on the temperature T of the correlation maximum is always a small fraction of the 3 dB linewidth of the main correlation function peak. In the experiments described here, the final temperature result found by correlation was never more than a few hundred millikelvin from the nominally correct result even when the line widths and relative TE and TM mode powers of reference spectrum $S_R$ (T, $\lambda$) were set several times larger or smaller than the observed values. This accuracy of less than 1 K is more than sufficient for most practical thermometer applications in industrial processes and environment monitoring.

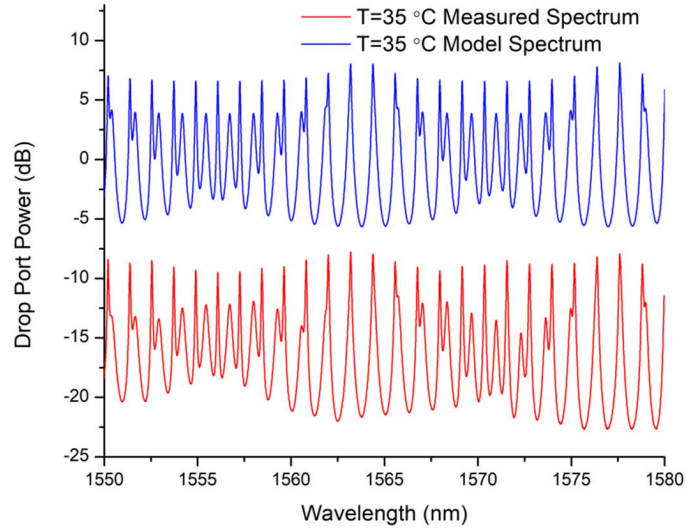

Fig.8. Comparison of model and measured ring spectra at T=35 °C. To generate the model spectrum the ring resonator coupler through/cross ratios are set at 0.8/0.2 (TE polarization) and 0.55/0.45 (TM polarization). The model TE and TM mode power ratio is adjusted to be in a 2:1 ratio. Spectra are offset for clarity.

To assess the correlation-based transduction procedure for thermometry, ring temperatures were determined from the maximum of the correlation function $G(T)$ in Eq. 1 over temperature T, for the original measured spectra over the range $\lambda_1$=1550 to $\lambda_2$ = 1580 and T = 20 and 50 °C shown in Figure 4. The temperatures determined from pattern recognition and the nominally correct temperatures (as measured by the stage thermistor) agree to within ±50 mK. These offsets are included in Figure 9 (see T scan data). The same temperatures were returned by spectral correlation when the sampled correlation scan temperature interval was only a few degrees or several hundred degrees wide, illustrating that the resonance pattern has a unique one-to-one correspondence with temperature over as very wide range.

In the case of the one-dimensional (i.e. temperature only) correlation search described above, the wavelength calibration of the laser source is implicitly assumed to be correct. However, since temperature is determined from a spectral pattern rather than the absolute wavelength of any specific resonance, it is possible to extract a temperature measurement from the ring spectrum even when the wavelength scale of light source or spectrometer used to interrogate the ring is not accurately calibrated. As a simple illustration, the correlation function of Eq. 1 was modified so that both temperature and an unknown constant wavelength calibration shift $\Delta\lambda$ are free parameters of the correlation search algorithm. In other words, we allow for the possibility that the laser wavelength readings may have become shifted by $\Delta\lambda$ from the true wavelength. The corresponding correlation function becomes

$$G(T, \Delta\lambda) = A \left| \int_{\lambda_1}^{\lambda_2} (S_M - \langle S_M \rangle)(S_R(T, \lambda + \Delta\lambda) - \langle S_R(T, \lambda + \Delta\lambda) \rangle) d\lambda \right|^2. \qquad (4)$$

Here the reference spectrum $S_R(T, \lambda+\Delta\lambda)$ is now a function of the trial temperature T and a trial wavelength calibration shift $\Delta\lambda$. By scanning T and $\Delta\lambda$ to find the maximum of the correlation function $G(T, \lambda+\Delta\lambda))$ over the (T, $\Delta\lambda$)-plane, both the ring temperature and a wavelength calibration correction can be found simultaneously. For simplicity, the wavelength calibration error here was assumed to be a constant shift $\Delta\lambda$, but in practice wavelength calibration errors may vary across the spectral range of interest. However as long as the wavelength calibration error can be parameterized in a numerically manageable form, the same correlation algorithm can be easily extended to quantify the error and provide a wavelength calibration correction.

Figure 10 shows a series of correlation functions $G(T, \lambda+\Delta\lambda)$ for a measured spectrum at T= 35 °C, where the temperature and the wavelength error $\Delta\lambda$ are both scanned over a two-dimensional parameter space covering $\Delta\lambda$ = -300 pm to +300 pm and temperatures from T= 20 to 50 °C. Here again there is one-to-one correspondence between spectral pattern and a point in the (T, $\Delta\lambda$)-plane, and the global correlation function maximum occurs near the nominally correct values of both temperature (35 °C) and wavelength calibration shift ($\Delta\lambda$ = 0 pm). The same temperature and wavelength correction are obtained when the algorithm is set to scan up to several hundred degrees in temperature or tens of nanometers in wavelength shift. Secondary maxima of the correlation function do appear when the temperature is scanned over a wider range. These side lobes arise from the quasi-periodic nature of the TE and TM resonance spectra. For example, the correlation side lobes at the left and right in Figure 10 correspond to the temperature offset at which the reference spectrum TE modes are translated by one TE free spectral range relative to the measured spectrum. But since the TE and TM spectra and temperature induced shifts are incommensurate such correlation side lobes are always lower than the main correlation peak. Note that even in the unlikely circumstance that TE and TM resonance spacings are locally commensurate (i.e. have an integer ratio such as 1:2) over some wavelength interval, over a wide enough spectral window the spectral pattern never repeats exactly because of dispersion of the optical properties with wavelength and temperature. Figure 9 shows the temperature error and wavelength offset error for both the one and two-dimensional correlation scans of data sets taken for stage temperatures between T = 20 °C to 50 °C at 5 °C intervals. When the spectrum wavelength calibration is assumed to be correct and correlation algorithm only scans temperature (i.e. using $G(T)$ in Equation 1), the algorithm returns ring temperature to within ±50 mK of the nominally correct value, as noted previously. When the wavelength calibration shift $\Delta\lambda$ is also a free parameter (i.e. when using $G(T, \lambda+\Delta\lambda))$ in Equation 4), the retrieved temperature deviates more widely from the correct result. However most retrieved temperatures and wavelength calibration shift values are within 200 mK and 15 pm of the nominally correct targets, respectively. The temperature error is roughly proportional to the wavelength shift offset error returned by the correlation algorithm. The larger error in the two-dimensional correlation scan arises because the spectral pattern (i.e. relative peak positions) varies more slowly with wavelength than temperature, and hence the correlation search along the wavelength axis is more sensitive to systematic deviations of the measured spectrum from the model reference spectrum $S_R$ (T, λ). As noted previously this source of error will be reduced for rings with much narrower linewidths. As a final example, Figure 11 gives a comparison between the results of ring temperature measurements retrieved by one (T only) and two dimensional (T and Δλ) correlation as the ring is heated by the on-chip heater, with the stage fixed at 22 °C. Here the temperatures returned by both one-dimensional and two dimensional spectral correlation search agree to within 150 mK over the applied current range. The residual systematic difference between one and two dimensional correlations results is again due to the higher sensitivity of the two dimensional scan to differences between the model reference spectrum and the measured spectrum.

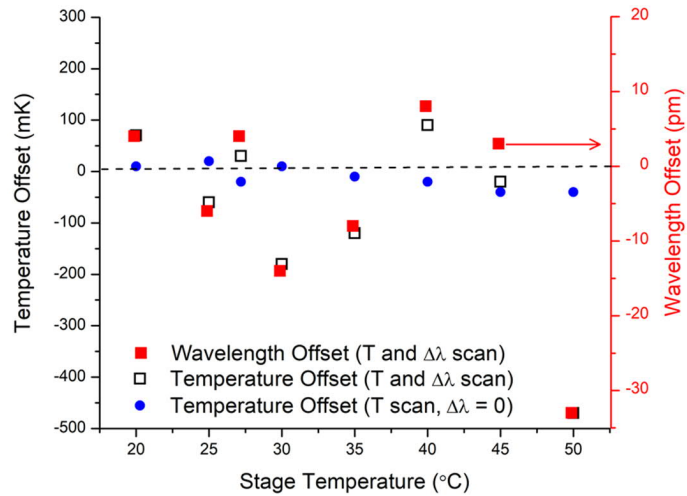

Fig.9. Offset in temperature T and wavelength calibration Δλ shift, as determined by two parameter (T, Δλ) spectral correlation scans, relative to the nominally correct stage temperature and laser wavelength. Ring temperatures determined by one parameter correlation (scanning temperature only, assuming the wavelength calibration is correct) are also shown.

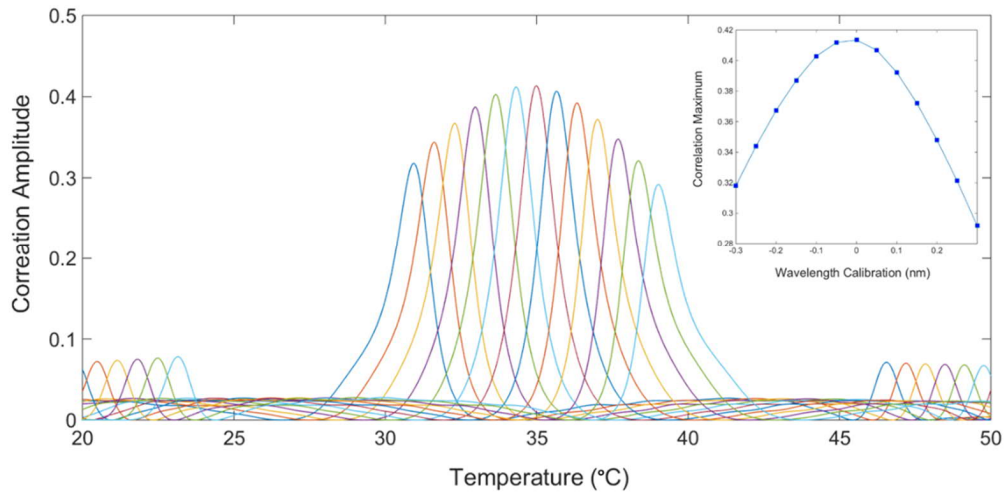

Fig.10. Correlation function $G(T,\lambda+\Delta\lambda)$ amplitudes over temperature for a measured ring spectrum at a stage temperature of 35 °C, for wavelength scale calibration shifts ranging from $\Delta\lambda$ = -300 to +300 pm where $\Delta\lambda$ = 0 pm is the nominally correct shift. The inset shows the correlation maximum as a function of wavelength calibration shift.

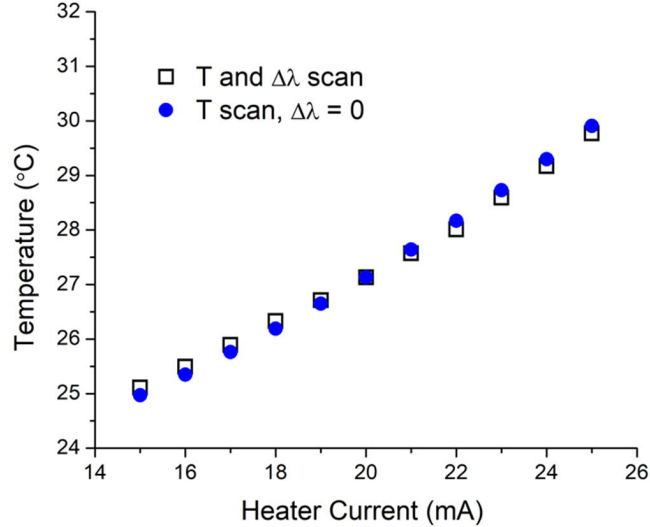

Fig.11. Variation of local ring temperature with micro-heater current as determined using spectral pattern recognition using one parameter $G(T)$ and two parameter $G(T, \lambda+\Delta\lambda)$ correlation functions.

## 5. Conclusions

The accuracy and resolution of pattern recognition transduction is determined by the linewidths of the ring resonances. As the correlation peak widths (see Fig. 10) broaden in proportion to increasing resonance width, the temperature obtained from correlation maximum becomes more susceptible to errors caused by deviations of the model and measured spectra, instrument related distortions and noise in the measured spectra. The systematic errors can in principal be reduced by using a more elaborate reference spectrum model in the correlation algorithm, but this will not correct for measurement system distortions arising from optical reflections, wavelength dependent coupling losses, and the shape of the incident light source spectrum. To approach the ultimate goal of a system independent temperature and wavelength metrology standard, an optimized ring should have sufficiently narrow resonances so that exact resonance wavelengths and hence temperature can be determined to the desired accuracy, independent of intrinsic line shapes and external distortions. In this work, the uncertainty in temperature measurement of 50 mK as in Fig. 9 is approximately 5% of the correlation half width of 1.3 °C shown in Fig. 10. The current resonance FWHM widths are 80 pm and 400 pm for TE and TM modes respectively. Assuming the accuracy of the correlation search scales linearly with linewidths, a ring with TE and TM resonance line FWHM less than 10 pm should approach temperature measurement uncertainties approaching 1 mK. Note that ring resonators designed with low loss waveguides have measured linewidths as small as 0.1 pm [23]. Similarly the precision of the wavelength combs obtained by locking to a mode coincidence point should scale to less than 1 pm (0.01 Å). Since in this work the ring linewidths are largely determined by ring coupling ratios, and reducing the linewidth to 10 pm can be addressed by optimizing coupler design, albeit with the added design complexity of optimizing both the TE and TM coupling ratios simultaneously. Increased accuracy and temperature resolution may also be improved by increasing the difference between the TE and TM waveguide thermo-optic responses, so that the pattern evolves more quickly with temperature change. This could be achieved by using a thinner Si waveguide, but such a design may not be fully compatible with the standard Si waveguide platforms and fabrication. Although ring design can clearly be improved, the temperature measurement accuracy of the present design is already sufficient for most practical applications. Ring temperatures have been determined to within 50 mK of the nominally correct values when the system wavelength calibration is assumed to be correct. When the wavelength calibration error is treated as an unknown variable, the correlation method was able to extract a temperature and a wavelength calibration correction simultaneously, to an accuracy of 200 mK and 15 pm, respectively.

In summary, pattern recognition is used to make temperature measurements using a single ring resonator with incommensurate TE and TM polarized resonances that in combination produce spectral patterns that are unique for every temperature. In effect each pattern is like a barcode that represents a specific temperature. Once the temperature of the ring is established from the spectral pattern, the same ring output spectrum can be used as a wavelength reference comb for calibration of spectroscopic instruments and light sources, without the need for separate measurement of the ring temperature. The spectral pattern is not cyclic so the temperature measurement range is limited only to the safe operating range of the chip and its packaging. Pattern recognition can also be intrinsically tolerant of wavelength calibration errors in the interrogating instruments, since temperature is encoded in the relative positions of peaks rather than their absolute wavelengths. Every spectral pattern represents a unique object in a two-dimensional temperature-wavelength space. If the pattern can be identified, the ring temperature and the wavelength of every ring resonance are then known (based on the initial calibration of the ring). Thus pattern recognition can in fact be used to quantify and correct wavelength calibration errors at the same time as temperature is retrieved, as we have shown experimentally. As a result, this method of transduction may be particularly interesting as a means to reduce the complexity and cost of the systems used to interrogate the ring. For example in the real time calibration of spectrometers often required in astronomy, a spectral pattern from the incommensurate ring can be coupled into the spectrometer, either in parallel or alternating with the science spectrum of interest. The ring pattern recorded by the spectrometer can be analyzed using a simple correlation algorithm and the temperature and absolute wavelength of every ring line will then be known, providing a direct calibration for the spectrometer. The light source can be a broadband source or low cost tunable laser, and there is actually no requirement to precisely measure or control ring temperature by other means. Similarly, for optical thermometry applications the wavelength calibration requirement for the interrogating light source or spectrometer can be relaxed, since pattern analysis can provide the wavelength calibration and an accurate temperature. The correlation algorithms are very simple so measurement and pattern analysis can be carried out essentially in real time with very modest computation resources. The focus of this work has been on temperature measurement, but similar pattern based transduction strategies may be applied to other silicon waveguide sensors, e.g. [24], such that the spectral pattern at any instant in time gives complete information on the state of the sensor.

**Disclosures**

The authors declare no conflicts of interest.

# References

1. A. Yariv, "Universal relations for coupling of optical power between microresonators and dielectric waveguides" Electronics Lett. **36** (4), 321-322 (2000)
2. W. Bogaerts, P. De Heyn, T. Van Vaerenbergh, K. De Vos, S. Kumar Selvaraja, T. Claes, P. Dumon, P. Bienstman, V. Van Thourhout, and R. Baets. "Silicon microring resonators," Laser and Phot. Rev. **6,** 47-73 (2011)**.**
3. G.-D. Kim, H.-S. Lee, C.-H.Park, S.-S. Lee, B. T. Lim, H.K. Bae, and W.-G. Lee, "Silicon photonic temperature sensor employing a ring resonator manufactured using a standard CMOS process," Opt. Express **18** (21), 22215-22221 (2010).
4. H.-T. Kim and M. Yu, "Cascaded ring resonator-based temperature sensor with simultaneously enhanced sensitivity and range," Opt. Express **24,** (9) 9501-9510 (2016).
5. D.-X. Xu, A. Delâge, P. Verly, S. Janz, S. Wang, M. Vachon, P. Ma, J. Lapointe, D. Melati, P. Cheben and J.H. Schmid, "Empirical model for the temperature dependence of silicon refractive index from O to C band based on waveguide measurements," Opt. Express **27** (19) 27229-27242 (2019).
6. H. Xu, M. Hafezi J. Fan, J.M.Taylor, G.F. Strouse and Z. Ahmed, "Ultra-sensitive chip-based photonic temperature sensor using ring resonator structures," Opt. Express 22 3098-3104 (2014).
7. S. Dedyulin, A. Todd, S. Janz, D.-X. Xu, S. Wang, M. Vachon, J. Weber, "Packaging and precision testing of fiber Bragg grating and silicon ring resonator based thermometers: current status and challenges," Measurement Sci. and Tech., in press, https://doi.org/10.1088/1361-6501/ab7611 (2020).
8. N. Klimov, T. Purdy, Z. Ahmed, "Towards replacing resistance thermometry with photonic thermometry," Sens. and Actuators A: Physical 269, 308–312 (2018).
9. P. R. N. Childs, J. R. Greenwood, and C. A. Long, "Review of temperature measurement," Review of Scientific Instruments **71**, 2959-2978 (2000).
10. J.V. Nicholas, D.R. White "Traceable Temperatures" (Wiley, 1995), Chapter 5, "Platinum Resistance Thermometry", pp. 153-198.
11. F. Kerber; F. Saitta; P. Bristow; J. Vernet, "Wavelength calibration sources for the near infrared arm of X-shooter," Proc. SPIE 7014, Ground-based and Airborne Instrumentation for Astronomy II, 129-141 (2008).
12. M. T. Murphy, P. Tzanavaris, J. K. Webb, C. Lovis, "Selection of ThAr lines for wavelength calibration of echelle spectra and implications for variations in the fine-structure constant" Monthly Notices of the Royal Astronomical Society **378** (1), 221–230 (2007).
13. R.A. Mccracken. J.M. Charsley and D.T. Reid, "A decade of astrocombs: recent advances infrequency combs for astronomy," Opt. Express **25** (13), 15058-15078 (2017).
14. R. A. McCracken, É. Depagne, R. B. Kuhn, N. Erasmus, L. A. Crause, and D. T. Reid, "Wavelength calibration of a high resolution spectrograph with a partially stabilized 15-GHz astrocomb from 550 to 890 nm," Opt. Express 25(6), 6450–6460 (2017).
15. M.-G. Suh, X. Yi, Y.-H. Lai, S. Leifer, I.S. Grudinin, G. Vasisht, E.C. Martin, M.P. Fitzgerald, G. Doppmann, J. Wang, D. Mawet, S.B. Papp, S.A. Diddams, C. Beichman and K.Vahala, "Searching for exoplanets using a microresonator astrocomb," Nature Photonics **13,** 25–30, (2019).|
16. F. F. Bauer, M. Zechmeister, and A. Reiners "Calibrating echelle spectrographs with Fabry-Pérot etalons," Astron. and Astrophys. **581**, A117 (2015).
17. C. Lee, S.T. Chu, B.E. Little, J. Bland-Hawthorn and S. Leon-Saval, "Portable frequency combs for optical frequency metrology," Opt. Express **20** (15), 1667-1676 (2012).
18. C M. Ibrahim, J. H. Schmid, A.Aleali, P. Cheben, J. Lapointe, S. Janz, P.J. Bock, A. Densmore, B. Lamontagne, R. Ma, D.-X. Xu, and W. N. Ye, "Athermal silicon waveguides with bridged subwavelength gratings for TE and TM polarizations," Opt. Express 20 (16), 18356-18361 (2012).
19. P. Cheben, J.H. Schmid, S.Wang, D.-X. Xu, M. Vachon, S. Janz, J. Lapointe, Y. Painchaud, and M.-J. Picard "Broadband polarization independent nanophotonic coupler for silicon waveguides with ultra-high efficiency." Opt. Express **23 (**17) 22554-22563 (2015)
20. C.Y. Cengel and A. Ghajar, Heat and Mass Transfer: Fundamentals and Applications, 5th edition (McGraw-Hill, 2015).
21. C S.M. Simpkin, Measurements of velocity dispersions and Doppler shifts from digitized optical spectra Astron and Astrophys 31, 129-136 (1974).
22. M. Brogi1, I. A. G. Snellen1, R. J. de Kok, S. Albrecht, J. L. Birkby, and E. J. W. de Mooij "Detection of molecular absorption in the dayside of exoplanet 51 pegasi," Astrophysical J. **767**, 27-37 (2013).
23. A. Biberman, M.J. Shaw, E.Timurdogan, J.B. Wright, and M.R. Watts. "Ultralow-loss silicon ring resonators," Opt. Lett. 37 (20), 4236-4238, 2012.
24. S. Janz, D.-X. Xu, M. Vachon, N. Sabourin, P. Cheben, H. McIntosh, H. Ding, S. Wang, J. H. Schmid, A. Delâge, J. Lapointe, A. Densmore, R. Ma, W. Sinclair, S.M. Logan, R. MacKenzie, Q.Y. Liu, D. Zhang, G. Lopinski, O. Mozenson, M. Gilmour, and H. Tabor, "Photonic wire biosensor microarray chip and instrumentation with application to serotyping of Escherichia coli isolates," Optics Exp. 21 (4), 4623-4637 (2013).